\documentclass[a4paper,11pt]{article}

\usepackage{jheppub}
\usepackage{graphicx,tabularx}
\usepackage{amsmath, amssymb, amsthm, amsfonts}
\usepackage[dvipsnames]{xcolor}
\usepackage{soul}
\usepackage{comment}
\usepackage{slashed}
\usepackage[yyyymmdd,hhmmss]{datetime}
\usepackage[backgroundcolor=white, bordercolor=white, textsize=tiny]{todonotes}
 
\usepackage{pbox}
\usepackage{multirow}
\usepackage{tikz-cd}
 
\newcommand{\tev}{\text{TeV}}
\newcommand{\gev}{\text{GeV}}
\newcommand{\ifb}{\text{fb}^{-1}}

\newcommand{\pb}{\text{ pb}}
\newcommand{\fb}{\text{ fb}}
\newcommand{\Br}{\text{ Br}}

\newcommand{\mh}{m_h}
\newcommand{\mH}{m_H}
\newcommand{\mA}{m_A}
\newcommand{\mC}{m_{H^\pm}}
\newcommand{\mx}{m_{12}^2}

\newcommand{\bpa}{\textbf{BP-A}}
\newcommand{\bpb}{\textbf{BP-B}}

\renewcommand{\eqref}[1]{Eq.~(\ref{#1})}

\newcommand{\figref}[1]{Fig.~\ref{fig:#1}}

\newcommand{\be}{\begin{equation}\begin{aligned}}
\newcommand{\ee}{\end{aligned}\end{equation}}

\RequirePackage[normalem]{ulem}

\newcommand{\GeV}{\textrm{ GeV}}
\newcommand{\TeV}{\textrm{ TeV}}
\newcommand{\Lu}{\mathcal{L}}

\title{Probing Exotic Charged Higgs Decays in the Type-II 2HDM through Top Rich Signal at a Future 100 TeV $pp$  Collider}
\author[a]{Shuailong Li,}
\author[a]{Huayang Song,}
\author[a]{Shufang Su}

\affiliation[a]{
  Department of Physics,
  University of Arizona,
  Tucson, Arizona 85721, USA
}

\emailAdd{shuailongli@email.arizona.edu}
\emailAdd{huayangs@email.arizona.edu}
\emailAdd{shufang@email.arizona.edu}

\preprint{
\begin{flushright}
\end{flushright}
}

\abstract{The exotic decay modes of non-Standard Model  Higgs bosons are efficient in probing the hierarchical   Two Higgs Doublet Models (2HDM). In particular, the decay mode $H^\pm\to HW^\pm$ serves as a powerful channel in searching for   charged Higgses.  In this paper, we analyzed the reach for $H^\pm\to HW^\pm \to t\bar{t}W$ at a 100 TeV $pp$ collider, and showed that it extends the reach of the previously studied $\tau\tau W$ final states once above the top threshold.  Top tagging technique is used, in combination with the boosted decision tree classifier.   Almost the entire hierarchical Type-II 2HDM parameter space can be probed via the combination of all channels at low $\tan\beta$ region.}

\begin{document} 

\titlepage
\maketitle

\section{Introduction}
The discovery of the 125 GeV Standard Model (SM)-like Higgs boson at the Large Hardon Collider (LHC) \cite{Aad:2012tfa,Chatrchyan:2012xdj} marks the great triumph of the SM in particle physics. The determination of its mass as well as spin has been well established~\cite{Aad_2013,CMS-PAS-HIG-19-004}.  Subsequent measurements of its couplings to SM particles are consistent with the SM predictions~\cite{Sirunyan_2019, Aad:2019mbh}.  There are, however, unsolved puzzles at both theoretical and experimental fronts, such as the hierarchy problem, neutrino mass and   nature of dark matter, which motivate physicists to explore new physics beyond the SM (BSM). Most of those BSM models involve an extended   Higgs sector, with two Higgs Doublets being one of the simplest options. 

In addition to the SM-like Higgs boson, the   spectrum of the Higgs sector in the two Higgs Doublet Model (2HDM) after electroweak symmetry breaking (EWSB) contains four   non-SM Higgses, the neutral CP-even Higgs $H$, the neutral CP-odd Higgs $A$ and a pair of  charged Higgses $H^\pm$. Two general experimental methods  are used to search for non-SM Higgses in the 2HDM.   One is indirect search using   electroweak precision measurements and  Higgs coupling measurements. The other is   direct search  of new particles at high energy colliders. For indirect search, the constraints of the latest LHC measurements on the 2HDMs parameter space are studied in~\cite{Haller_2018} while the implications of the future $Z$ and Higgs factory precision measurements on the Type-I and Type-II 2HDMs are explored in~\cite{chen2019typei, Chen_2019typeii, Gu:2017ckc}.  While the allowed regions are   reduced with the improved precision if no deviation from the SM prediction is observed, certain parameter space is hard to reach, for example, the alignment limit of the 2HDM.  In contrast, direct search at colliders is very sensitive to the masses and couplings of new particles.  In this work, we focus on the direct search of the charged Higgses in the Type-II 2HDM. 

The direct search of non-SM Higgs bosons mostly focuses on their decays into other SM particles, similar to the channels of the 125 GeV SM-like Higgs boson study at the LHC. However, these modes are limited in the search for the non-SM Higgses. In general, the fermionic decay suffers from either small branching fractions or large SM backgrounds.  Decays to $WW$ and $ZZ$  are   suppressed given  that the observed 125 GeV Higgs is very SM-like.   Fortunately,  the couplings among two non-SM Higgses and one SM gauge boson or among three Higgses   are unsuppressed.  Once kinematically allowed, the exotic decays of a heavy non-SM Higgs into a light non-SM Higgs plus one SM gauge boson, or into two light non-SM Higgses dominate.  Discovery potential of such exotic decay modes of non-SM like Higgses has been explored in the literature for  $A/H\to HZ, AZ$~\cite{Coleppa_2014}, $H/A\to H^\pm W^\mp$~\cite{Li_2015} and $H^\pm \to AW^\pm/HW^\pm$~\cite{Coleppa_2014_charged, Patrick_2018, Kling:2015uba} at the 14 TeV LHC. The searches of $A/H\to HZ/AZ$ channels have been carried out at both ATLAS~\cite{Aaboud_2018_neutral_Z} and CMS~\cite{Khachatryan_2016} experiments.    The reach for exotic channels  at a future 100 TeV $pp$ collider are recently studied as well~\cite{Kling_2019}.

Among all the non-SM Higgs bosons, the search of charged Higgses is  challenging, especially for $m_{H^\pm}>m_t$.   For $m_{H^\pm}<m_t$, the charged Higgs can be produced in top decay, and  primarily decays into $\tau\nu$ and $cs$.   The null search results exclude all values of $\tan\beta$ for $m_{H^\pm}\lesssim 160\GeV$ in the Type-II 2HDM~\cite{aleph2013search,Aaboud_2018_charged, Sirunyan_2019_hpm}. For $m_{H^\pm}>m_t$, the dominant production mode is $tbH^\pm$, with relatively small production cross sections, compared to those  of the neutral Higgses. The  dominant decay mode of $H^\pm \to tb$, unfortunately, has   overwhelming large SM backgrounds. In the type-II 2HDM, the current LHC exclusion limit is weak~\cite{Aaboud_2018_tb}:  only regions of $\tan\beta>40$ and $<2$ at $m_{H^\pm}\sim 300\GeV$ are excluded. The reach will be further reduced once the exotic decay mode of $H^\pm \to AW^\pm/HW^\pm$ opens up. 

The exotic decay mode of $H^\pm \to AW^\pm/HW^\pm$ provides an alternative channel for charged Higgs search. The reach of this channel via $A/H\to\tau\tau$ mode has been studied in literature \cite{Coleppa_2014_charged, Kling_2019}.    In the $m_{H^\pm}-\tan\beta$ plane, the exclusion region extends to $m_{H^\pm}=600\GeV$ in both small and large $\tan\beta$ regimes at the 14 TeV LHC with 300 $\fb^{-1}$ luminosity~\cite{Coleppa_2014_charged}.  At a  future 100 TeV $pp$ collider, the reach is further extended to 2.5 TeV for $\tan\beta \gtrsim 10$~\cite{Kling_2019}.  While  $\tau\tau$ channel is effective  due to its clean signature, its sensitivity is reduced once the  heavy neutral Higgs mass is above the top pair threshold or at small $\tan\beta$.  The $A/H\to t \bar t$ channel has been studied in literature~\cite{Patrick_2018} focusing on the signal events with trilepton and same-sign dilepton (SSDL) signature at the 14 TeV LHC with 30 $\fb^{-1}$ luminosity. It shows great sensitivity in the $m_{H/A} - m_{H^\pm}$ plane at $m_{H^\pm}>500$ GeV.  In particular, for low $\tan\beta \sim 1$, $t\bar{t}$ channel extends the reach of charged Higgs to about 1 TeV, well beyond the reach of $\tau\tau$ mode. 

In recent years, there have been worldwide  efforts of proposing   a 100 TeV $pp$ collider, including the Future Circular Collider (FCC) at CERN \cite{fccplan} and the Super proton-proton Collider (SppC) in China \cite{CEPC-SPPCStudyGroup:2015csa}.   Given the high energy environment of such machine, highly boosted object such as top quark can be identified using top-tagging technique~\cite{Plehn_2010,Plehn_2012,Kling_2012,Kaplan_2008,Thaler_2012,Kasieczka_2017}, which offers additional handle for new physics discovery by suppressing SM hadronic backgrounds.    This  technique has been implemented in searching the exotic decay of $A\to HZ$, with $H\to t\bar{t}$ at the future 100 TeV $pp$-collier, which extends the reach of $H\to\tau\tau$ mode above the di-top threshold~\cite{Kling_2019}.  It's also widely used in conventional search modes such as $A/H\to t\bar t$ \cite{Guchait:2018nkp, Craig:2016ygr,Hashemi:2017ine}. 

In this study, we propose to search the charged Higgs via exotic decay   $H^\pm\to HW^\pm$ with $H\to t\bar t$ channel  at a future 100 TeV $pp$ collider, using top tagging technique.  We focus on the hierarchical scenario of $m_A=m_{H^\pm}>m_H$, which is motivated by theoretical constraints and electroweak precision measurement~\cite{Kling_2019}.  The Boosted Decision Tree (BDT) classifier~\cite{Chatrchyan:2012xdj} is used to distinguish the signal and background events.

The rest of the paper is organized as follows. In Sec.~\ref{sec:scenario} we briefly introduce Type-II 2HDM, as well as the charged Higgs interactions.  We also discuss benchmark plane for collider study and  the current experimental search bounds. In Sec.~\ref{sec:signal_analysis}, we describe the detailed collider analyses and the reach for the charged Higgs.  We concluded in Sec.~\ref{sec:conclusion}.

\section{Type-II 2HDM and Charged Higgses}
\label{sec:scenario}
Two ${\rm SU}(2)_L$ doublets $\Phi_i$, $i=1, 2$ are introduced in the 2HDM:
\begin{equation}\label{eq:higgs_doublets}
    \Phi_i=\left(\begin{matrix}
        \phi_i^+\\
        (v_i+\phi_i^0+i G_i)/\sqrt{2}
    \end{matrix}\right)
\end{equation}
where $v_1$ and $v_2$ are the vacuum expectation values (VEVs) of the neutral components which
satisfy the relation $v=\sqrt{v_1^2+v_2^2}=246\GeV$ after EWSB. In scenarios with only a soft breaking of a discrete $\mathcal{Z}_2$ symmetry allowed, the most general Higgs potential has eight parameters, which can be chosen as four physical Higgs masses $(\mh, \mH, \mA, \mC)$, the electroweak VEV $v$, the CP-even Higgs mixing angle $\alpha$, the ratio of the two VEVs, $\tan\beta=v_2/v_1$, and the $\mathcal{Z}_2$ soft breaking parameter $\mx$.

In our study of the heavy non-SM Higgses, we  assume the light netural CP even Higgs $h$ to be the observed SM-like $125\GeV$ Higgs. Current measurements on the properties of the $125\GeV$ Higgs  have already imposed strong constraints on the 2HDM parameter space~\cite{Haller_2018, Coleppa:2013dya},   pushing towards the \textit{alignment limit} of $\cos(\beta-\alpha)\simeq 0$.  Note that under   such limit, the charged Higgses $H^{\pm}$ preferably couple to the non-SM-like Higgses $A$ or $H$:
\begin{align}
    g_{H^{\pm}hW^{\mp}}&=\frac{g\cos(\beta-\alpha)}{2}(p_{h}-p_{H^{\pm}})^{\mu}\simeq 0, \\
    g_{H^{\pm}HW^{\mp}}&=\frac{g\sin(\beta-\alpha)}{2}(p_{h}-p_{H^{\pm}})^{\mu}\simeq\frac{g}{2}(p_{H}-p_{H^{\pm}})^{\mu}, \\
    g_{H^{\pm}AW^{\mp}}&=\frac{g}{2}(p_{A}-p_{H^{\pm}})^{\mu},
\end{align}
where $g$ is the ${\rm SU}(2)_L$ coupling and $p^{\mu}$ is the incoming momentum for the corresponding particle.    Once $m_{H^\pm}-m_{A/H}>m_W$, $H^{\pm}\rightarrow AW^{\pm}/HW^{\pm}$ opens up with sizable decay branching fractions.

We refer the readers to our previous papers (Ref.~\cite{Kling:2016opi, Kling_2019}) for more detailed discussion on the constraints on the 2HDM parameter space from theoretical considerations and electroweak precision measurements. For TeV-scale  masses, two benchmark planes are proposed for collider studies~\cite{Kling_2019}: \bpa~($m_{A}>m_{H}=m_{H^\pm}$) with $A\rightarrow HZ/H^\pm W^\mp$
and \bpb~($m_{A}=m_{H^\pm}>m_{H}$) with $A\rightarrow HZ$, $H^\pm\rightarrow H W^\pm$.  In this paper, we focus on the benchmark plane \bpb, which permits the decay of $H^{\pm}\rightarrow HW^{\pm}\to t\bar{t}W^{\pm}$.

Searches for charged Higgses have been performed both at the ATLAS and CMS experiments. A search with clean $\tau\nu$ final state by ATLAS using $36~\ifb$ integrated luminosity at $13~\tev$~\cite{Aaboud:2018gjj} excludes $H^{\pm}$ mass range up to $1100~\gev$ (400 GeV) at $\tan\beta=60$ (30) in the hMSSM scenario of the Minimal Supersymmetric Standard Model (MSSM).   The reach is reduced greatly at low $\tan\beta$ region though.  For $0.5<\tan\beta<10$, there is    only   a very weak lower bound for charged Higgs mass ($\mC\gtrsim 160~\gev$). The CMS results~\cite{Sirunyan:2019hkq} are similar. 

Both ATLAS and CMS have also searched for a heavy charged Higgs boson produced in association with a top quark with $H^{\pm}\rightarrow tb$~\cite{Aaboud:2018cwk, Sirunyan:2019arl, Sirunyan:2020hwv}, which is sensitive to both the small and large $\tan\beta$ region.   The null search results at ATLAS using multi-jet final states with one or two electrons or muons~\cite{Aaboud:2018cwk} impose an upper limit for $\sigma(pp\rightarrow H^{\pm}tb)\Br(H^{\pm}\rightarrow tb)$ of $0.07\pb$ at $\mC=2000~\gev$ and $2.9\pb$ at $\mC=200~\gev$.  When interpreted in the hMSSM scenario, the charged Higgs with $\mC$ up to 200 (965) $\gev$ for $\tan\beta\sim 1.95 (0.5)$ has already been ruled out.  The CMS results~\cite{Sirunyan:2019arl} using events with a single isolated electron or muon or an opposite-sign electron or muon pair are slightly better, while the limits with all-jet final states are weaker~\cite{Sirunyan:2020hwv}.

Various flavor measurements~\cite{Amhis:2016xyh, Haller:2018nnx}, mostly from $B$-system, also provide indirect constraints on the charged Higgs mass $\mC$. The most stringent of these comes from the measurement of the branching fraction of the
decays $b\rightarrow s\gamma$ and $B^+\rightarrow\tau\nu$, which conservatively disfavor $\mC<580~\gev$ in the Type-II 2HDM~\cite{Misiak:2017bgg}.  Flavor constraints, however, can be relaxed with contributions from other sectors of new physics models~\cite{Han:2013mga}. In the following study,  we focus on the direct collider reach of charged Higgses.

Note that direct searches for non-SM neutral Higgs also provide constraints on the 2HDM parameter space.  We refer   interested readers to Ref.~\cite{Kling:2016opi, Kling_2019} for details. A recent paper~\cite{Kling:2020hmi}  summarises the status of neutral Higgs searches at the LHC, with both direct and indirect measurements.

\section{Signal Analysis}\label{sec:signal_analysis}
 In this section we perform the collider analyses of charged Higgs with $tbH^\pm$ associated production  with the  subsequent decay of $H^{\pm}\rightarrow HW^{\pm}\to t \bar{t}W^{\pm}$ at a 100 TeV $pp$ collider in  the $m_A=m_{H^\pm}>m_H$ parameter region (\bpb). We require the final state containing two hadronically decaying top, and a pair of charged leptons.

\subsection{Charged Higgs Production   and Decay} 
The dominant production  channel for heavy charged Higgses at a  $pp$ collider is $gg\to H^\pm tb $. The production cross section calculated at next to leading order (NLO) using Prospino \cite{beenakker1996prospino,Plehn_2003} is plotted in the left panel of Fig.~\ref{fig:branchratio}. The suppression at $\tan\beta\sim 7$ is due to  the suppression of the $H^\pm tb$ coupling at intermediate value of $\tan\beta \sim \sqrt{m_t/m_b}$.

The decay branching fractions of $H^\pm$ as a function of their mass are shown in the right panel of Fig.~\ref{fig:branchratio} for $m_H=600$ GeV and $\tan\beta=1.5$.  For $m_{H^\pm}<m_H+m_W$, $H^\pm\to tb$ dominates, with $H^\pm \to \tau\nu, cs$ being almost negligible.    However, once $m_{H^\pm}>m_H+m_W$,  $H^{\pm}\rightarrow HW^{\pm}$ quickly dominates, as shown by the black curve.   All branching fractions are calculated using 2HDMC~\cite{Eriksson_2010}.

\begin{figure}[h]
\centering
\includegraphics[width=0.45\textwidth]{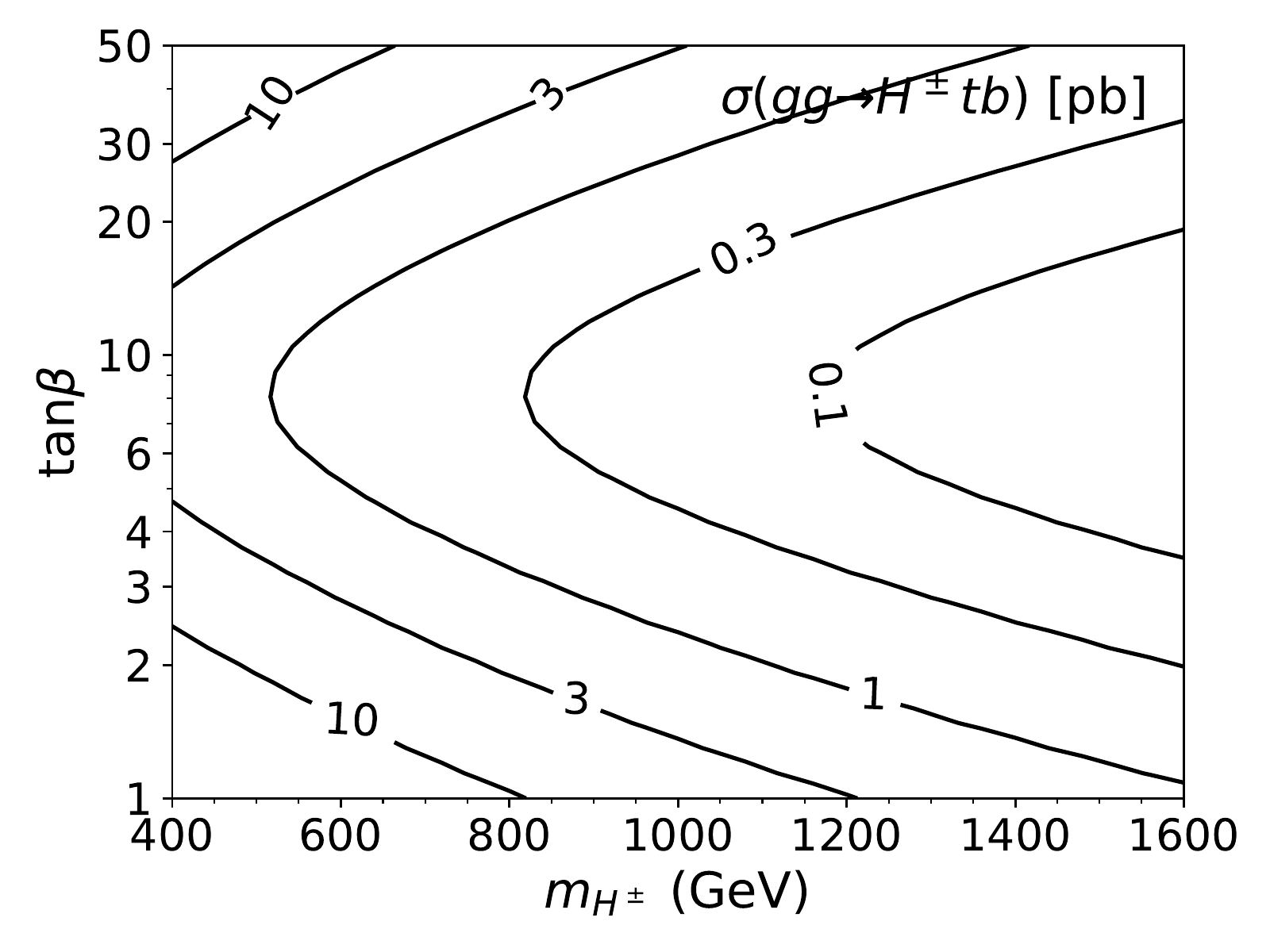}
\includegraphics[width=0.45\textwidth]{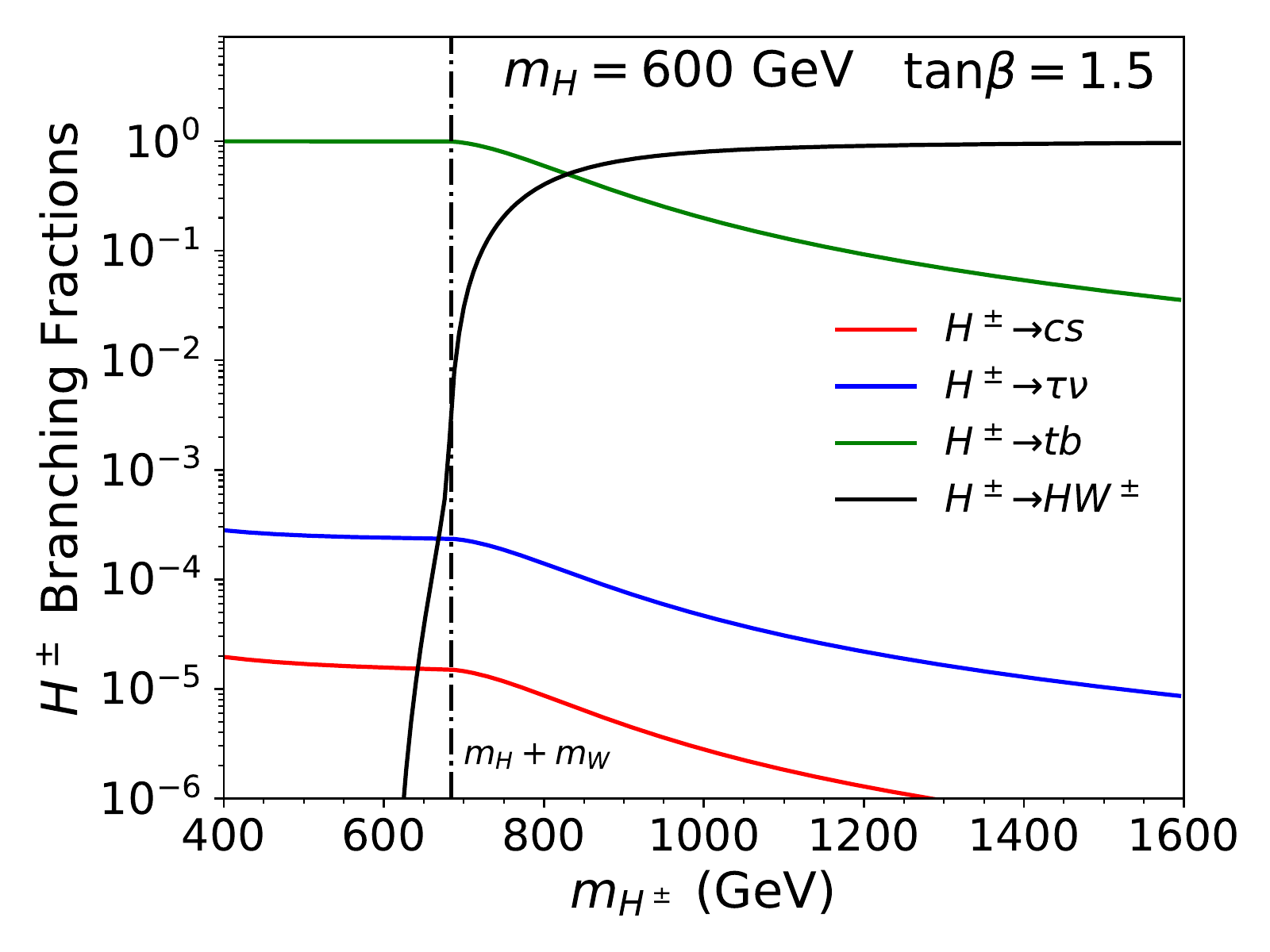}
\caption{The left panel shows the charged Higgs production cross section of  $gg\to H^\pm tb$ in the  $m_{H^\pm}$ vs. $\tan\beta$ plane for a 100 TeV $pp$ collider.   The right panel shows the branching fractions of charged Higgses for  $H^\pm\to cs$ (red), $\tau\nu$ (blue), $tb$ (green), $HW^\pm$ (black) at $\tan\beta=1.5$, $m_{H}=600\GeV$. The dash-dotted lines indicates the mass threshold  at $m_H+m_W$.    }
\label{fig:branchratio}
\end{figure}

\subsection{Search Strategy}

The exotic decay of $H^\pm\to HW^\pm$  has been   studied via $H\to\tau\tau$ channel~\cite{Kling_2019}, with same-sign dilepton events by requiring one $\tau$ and one $W$ of the same sign decaying leptonically while the other $\tau$ and $W$ decaying hadronically.  The main background  comes from $t_ht_\ell(Z/h/\gamma^*\to \tau_h\tau_\ell)$.  While this search mode has a good reach at large $\tan\beta$, the reach is limited at low $\tan\beta$, especially for $m_H>2 m_t$ when $H\to t\bar{t}$ dominates.  

For these reasons, in this study, we explore the reach of $H\to t\bar t$ channel for charged Higgs discovery.  In particular, top tagging technique can be used to identify the highly boosted top produced from heavy Higgs decay.  The production and decay chain is
\begin{equation}\label{eq:Hpm_decaychain}
gg\to tbH^\pm\to tbHW^\pm\to t\bar ttbW^\pm
\end{equation}
With the subsequent decay of $t\to bW$, there are four $W$s in the final state.   We require two of them decay hadronically, which are needed for top-tagging or top reconstruction,  and the other two decay leptonically.  The event signature for detection is $\ell\ell+b+2t_h+\slashed{E}_T$, where $\ell=e,\ \mu$.  Note that for 2 $t_h$, we mean at least one $t_h$ being top-tagged, while the other being either top-tagged or reconstructed, as explained in details below.  Extra $b$-jet is required here to suppress the SM $ttZ$ background.   

The dominant irreducible SM background arises from four-top process with two tops decaying leptonically and the other two decaying hadronically.  The total cross section of the four-top production at $100\TeV$ $pp$ collider   calculated using MadGraph 5 \cite{Alwall_2014} at leading order (LO) is $\sigma(gg\to t\bar tt\bar t)=2.97\pb$.  We use a $K$-factor of 1.66 to take into account the NLO effect~\cite{Mangano:2016jyj}. The subdominant background is the $tttbW^\pm$ process with $bW^\pm$ in the final state not coming from a resonant top.  The corresponding LO cross section is $\sigma(gg\to tttbW)=0.623\pb$.  

As mentioned earlier, the top quarks originating from the neutral Higgs $H$ are highly boosted for a heavy   Higgs $H$, with top decay products grouping into a cone of size $\lesssim 1.5$. The top tagging techniques~\cite{Plehn_2010,Plehn_2012,Kling_2012,Kaplan_2008,Thaler_2012,Kasieczka_2017} developed in recent years could be used to identify the top quark in the signal process, while suppress both the SM hadronic backgrounds and the SM background with softer top quarks. However, due to the complexity of the  final states, the hadronic $W$ boson from $H^\pm \to HW^\pm$ along with the associated bottom quark have non-negligible probability to be identified as a hadronic top as well.   Furthermore, at a 100 TeV $pp$ collider, a considerable portion of tops produced associated with the charged Higgses are energetic enough to be captured by the top tagger.  Therefore, we need to consider all possible combinations of dilepton decay in Eq.~(\ref{eq:Hpm_decaychain}),  including same-sign dilepton (SSDL) as well as opposite-sign dilepton (OSDL) signals.  

To identify different combinations of top decays that lead to the dilepton signals, we list the sub-processes for SSDL and OSDL events in \autoref{tb:division}.
For OSDL, there are four sub-processes with OS1 being both hadronic decaying top from $H$ decay, OS2 and OS3 being one hadronic top from $H$ decay and one hadronic top from either the associated produced top, or the hadronic $W$ from $H^\pm$ decay,    and OS4 being the case when both tops from $H$ decay leptonically.  There are two sub-processes for SSDL, both of them having one hadronic top from $H$ decay, and the other one from either the associated top or the $W$.  

\begin{table}[h]
\centering
\begin{tabular}{ |c|c|c|@{}p{1mm}@{}|c|c|c| } 
\hline
\hline
    Channel & Divisions & Combinations & & Channel & Divisions & Combinations\\ 
\hline
\hline
\multirow{4}{3em}{OSDL} & OS1 & $(H\to t_h\bar t_h)t_{\bar \ell} \bar bW_\ell^-$ && \multirow{4}{2em}{SSDL} &&   \\ 
\cline{2-3}\cline{6-7}
 & OS2& $(H\to t_{\bar\ell}\bar t_h)t_h \bar bW_\ell^-$ &&& SS1 & $(H\to \bar t_{\ell}t_h)t_h \bar bW_\ell^-$\\ 
 \cline{2-3}\cline{6-7}
  & OS3& $(H\to t_h\bar t_{\ell})t_{\bar\ell} \bar bW_h^-$ &&& SS2& $( H\to \bar t_h t_{\bar\ell})t_{\bar \ell} \bar bW_h^-$\\ 
  \cline{2-3}\cline{6-7}
  & OS4& $(H\to t_{\bar\ell}\bar t_\ell)t_h \bar bW_h^-$ &&&&\\ 
 \hline
\end{tabular}
\caption{Divisions of the signal processes corresponding to different combinations of $W$ decays that lead to the dilepton signals in the final states.  }
\label{tb:division}
\end{table}

To perform a  detail  collider simulation, all signal and background events are generated using MadGraph5\_aMC@NLO v2.6.6~\cite{Alwall_2014}, then interfaced with Pythia8~\cite{Sj_strand_2015}  for showering and hadronization. The events are then passed into Delphes 3~\cite{de_Favereau_2014} for detector simulation using the Delphes card prepared by FCC-hh collaboration~\cite{FCC-Delphes-Card}. All branching fractions of Higgses and decay width are calculated by 2HDMC~\cite{Eriksson_2010}. 

For the purpose of triggering, we require the leading and sub-leading leptons to have the transverse momentum:  $p_{T,\ell_1}>20\GeV$ and  $p_{T,\ell_2}>10\GeV$. Other than the OSDL/SSDL signature, a $b$-tagged jet with $P_{T,b}>30$ GeV is also required.     In addition, we demand at least one hadronic top to be identified by top-tagging algorithm and a second hadronic top to be  either top-tagged or reconstructed for each event.  Specifically, for event with one tagged top, we further require it contains at least two bottom jets and at least two untagged jets. We then reconstruct another hadronic top following Ref. \cite{Guchait_2018} by iterating over all pairs of untagged jets and find the one that has invariant mass $|m_{jj}-m_W|\leq 20\GeV$.  We further combine the jet pair with $b$-jet that satisfies $|m_{bjj}-m_t| \leq  50\GeV$.  We also require the reconstructed top to have $P_{T,t}>100\GeV$.   Note that requiring two top tagging reduces the cut efficiency of the signal significantly due to the complexity of the final states.  Therefore we only require one tagged top   in the signal events.  

Next, we pass the selected events to a BDT classifier \cite{Thaler_2012} implemented in the Toolkit for Multivariate Data Analysis (TMVA)~\cite{hoecker2007tmva} to help distinguish the signal and backgrounds.   The observables used in BDT include
\begin{itemize}
\item the transverse momenta of the two hadronic  tops:  $p_{T,t_1}$ and $p_{T,t_2}$;
\item the transverse momenta of the leading and sub-leading lepton:  $p_{T,\ell_1}$ and $p_{T,\ell_2}$;
\item invariant mass of the two hadronic tops $m_{t_1,t_2}$;
\item the number of jets $N_j$;  
\item difference of pseudo-rapidity between the two leptons $\Delta \eta_{\ell_1\ell_2}$;
\item scalar sum of all transverse energy $H_T$ and miss transverse energy $\slashed{E}_T$.
\end{itemize}

We generate 200K events in each signal division   (1.2M total) as listed in Table.~\ref{tb:division} for a given benchmark point, 3M (1M) $t\bar{t}t\bar{t}$ events  for OSDL (SSDL) backgrounds, and 1M $tttbW$ events each for  OSDL and SSDL  backgrounds. We use half of the events to train the BDT and the other half to test. An optimal cut  on the BDT output is used to perform the hypothesis test at each benchmark point.  A minimum requirement of three events are imposed after the BDT cuts.   The projected statistical significance is obtained at an integrated luminosity $\Lu=3000\fb^{-1}$.   Assuming 10\% systematic error in the background cross sections, the expected exclusion and discovery significance $Z_{\mathrm{excl}}$ and $Z_{\mathrm{disc}}$ are estimated following Ref.~\cite{Kling_2019}. Note that two separate BDT cuts are chosen to maximize the discovery and exclusion significance separately.  We claim the search region has the potential to be discovered at 5 $\sigma$ significance if $Z_\text{disc}\geq5$ and excluded at 95\% C.L. if $Z_\text{excl}\geq 1.645$.  
 
\subsubsection{Opposite-Sign Dilepton Search}\label{sec:4_2_1}
 
To illustrate the cut efficiency and compare the significance between different OSDL signal divisions   and different benchmark points,  in  \autoref{tb:OSDL_search}, we show the cross sections after each step of cuts for two benchmark points at $\tan\beta=1.5$:  $m_{H^\pm}=1200$ GeV, $m_H=900$ GeV representing a heavy neutral Higgs and $m_{H^\pm}=800$ GeV, $m_H=600$ GeV representing a light neutral Higgs.  The combined background cross section for the dominant non-irreducible backgrounds of $tttt$ and $tttbW$ is given as well.   The other reducible background such as $tt(Z)$, $ttjj$ and $jjjj$  will be sufficiently suppressed after all levels of cuts,  comparing to the irreducible backgrounds of $tttt$ and $tttbW$.  The final $Z_\text{disc}$ and $Z_\text{exc}$ shown in the last column already include the 10\% systematic error.  
 
\begin{table}[h]
\centering
\begin{tabular}{ |c|c|c|c|c|c| } 
\hline
\hline
{\small\textbf{OSDL}}	 & {\small\textbf{Total [fb]}} & {\small\pbox{15cm}{\textbf{Dilepton +1 $b$+}\\ \textbf{1 tagged top [fb]}}} & {\small\textbf{Recon. [fb]}}& {\small\pbox{15cm}{\textbf{BDT [fb]}\\ \textbf{DISC/EXC}}} &{\small$\boldsymbol{Z_{\rm{disc}}/Z_{\rm{exc}}}$}\\
\hline
\hline
\multicolumn{6}{|c|}{$m_{H^\pm}=m_A=1200\GeV$, $m_H=900\GeV$}\\
 \hline
OS1 & 18.3 & 1.73	  & 0.377	 & - & -\\
OS2 & 18.3 & 1.58	  &0.292 & - & -\\
OS3 & 18.3 & 1.08	  &0.155 & - & -\\
OS4 & 18.3 & 0.650	&0.100 & - & -\\
\hline
 Combined & 73.08 & 5.03 &0.922 & 0.0789/0.111 & \multirow{2}{*}{ 3.79/4.71}\\
 \cline{1-5}
 Background & 471.5 &  31.6 & 8.17 & 0.165/0.240 & \\
\hline
\hline
\multicolumn{
6}{|c|}{$m_{H^\pm}=m_A=800\GeV$, $m_H=600\GeV$}\\
 \hline
OS1 & 39.3 & 1.92	 & 0.337& - & -\\
OS2 & 39.3 & 2.13 	& 0.392 & - & -\\
OS3 & 39.3 & 1.33	 &0.207& - & -\\
OS4 & 39.3 & 1.24	 &0.221 & - & -\\
\hline
 Combined & 157.2 & 6.62 &1.157 & 0.0365/0.117 & \multirow{2}{*}{6.65/8.73}\\
 \cline{1-5}
 Background & 471.5 & 31.6 & 8.17 & 0.0305/0.141 & \\
\hline
\end{tabular}
\caption{Cross sections of OSDL signal for two benchmark points and  dominant SM backgrounds after different steps of  cuts.   Also shown are the discovery and exclusion significance after BDT cuts for a 100 TeV $pp$ collider with 3000 ${\rm fb}^{-1}$ integrated luminosity.  }
\label{tb:OSDL_search}
\end{table}

For benchmark point $m_{H^\pm}=m_A=1200\GeV$, $m_H=900\GeV$, the signal events in  OS1 contribute the most to the signal cross section, consistent with the fact that the hadronic tops out of $H$ are energetic and relatively easier to be top tagged.     Note, however, that even in OS4 when the tagged top is the associated top, it still contribute to about 10\% of the total signal cross sections.   

As the transverse momenta of tops from $H$ decay reduce, the contribution from the hadronic associated top become   more prominent, as shown in benchmark $m_{H^\pm}=m_A=800\GeV$, $m_H=600\GeV$.  Signal events in OS2 contributes the most, while the contribution from OS4 is comparable to that of OS3.  This could be explained by the long tail in the $p_T$ distribution of the associated top towards high $p_T$ region, as comparing to that of the top from $H$ decay, as shown in the blue and red solid curves in Fig.~\ref{fig:OSDL_pt}, respectively.   Hadronic tops of higher $p_T$ are preferred by top tagger due to the higher tagging efficiency.  The dashed curves in Fig.~\ref{fig:OSDL_pt} shows the hadronic top $p_T$ distribution for benchmark $m_{H^\pm}=m_A=1200\GeV$, $m_H=900\GeV$ as a comparison.   The $p_T$ distribution of hadronic tops in OS3 is harder than that of OS4.   However, OS4 still permits a  considerable amount of energetic tops, which explain the   10\% contribution of OS4 to the final signal cross section. 

\begin{figure}[h]
\centering
\includegraphics[width=0.5\textwidth]{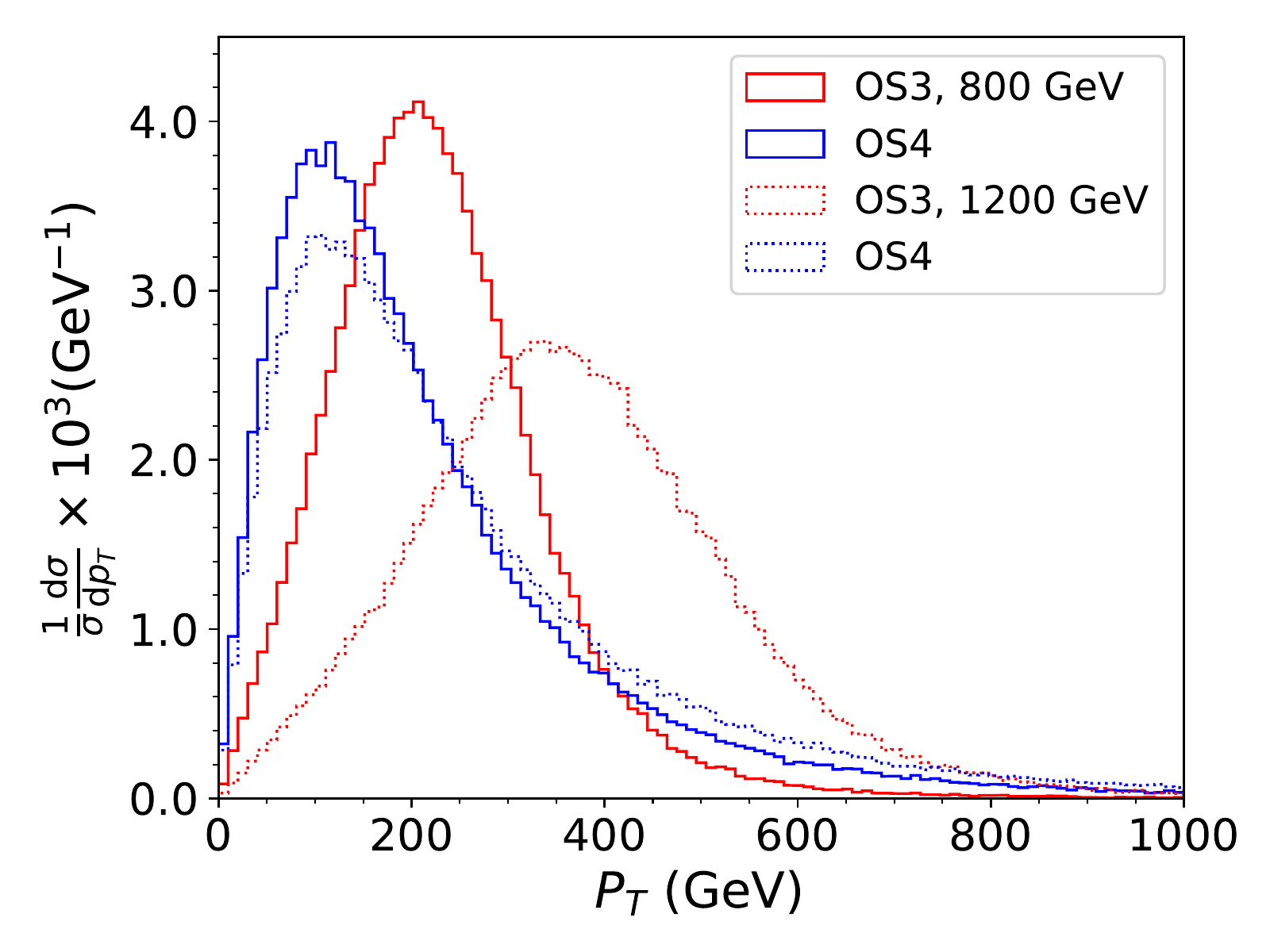}
\caption{The normalized $p_T$ distribution of the hadronic tops in signal divisions OS3 (red) and OS4 (blue) at parton level for benchmark points at $m_{H^\pm}=m_A=800\GeV$, $m_H=600\GeV$ (solid) and $m_{H^\pm}=m_A=1200
\GeV$, $m_H=900\GeV$ (dashed).    }
\label{fig:OSDL_pt}
\end{figure}

\subsubsection{Same-Sign Dilepton Search}
Given that considerable amount of tagged top coming from the associated top,  decays leading to SSDL would also contribute to the signal process.  The results are presented in \autoref{tb:SSDL_search}. Comparing  divisions SS1 and SS2 with divisions OS2 and OS3 in \autoref{tb:OSDL_search}, we find that the event rates are almost equal. The change of significance between OSDL and SSDL is primarily due to the difference in the total cross sections.    The combined OSDL and SSDL results are given in \autoref{tb:combined_search}. 

\begin{table}[h]
\centering
\begin{tabular}{ |c|c|c|c|c|c| } 
\hline
\hline
{\small\textbf{SSDL}}	 & {\small\textbf{Total [fb]}} & {\small\pbox{15cm}{\textbf{Dilepton +1 $b$+}\\ \textbf{1 tagged top [fb]}}} & {\small\textbf{Recon. [fb]}}& {\small\pbox{15cm}{\textbf{BDT [fb]}\\ \textbf{DISC/EXC}}} &{\small$\boldsymbol{Z_{\rm{disc}}/Z_{\rm{exc}}}$}\\
\hline
\hline
\multicolumn{6}{|c|}{$m_{H^\pm}=m_A=1200\GeV$, $m_H=900\GeV$}\\
\hline
SS1 & 18.3 & 1.61	  &0.303	 & - & -\\
SS2 & 18.3 & 1.09	  & 0.164	 & - & -\\
\hline
Combined & 36.6 & 2.70  &0.467 & 0.0525/0.0609 & \multirow{2}{*}{6.83/8.69}\\
\cline{1-5}
Background & 235.8 & 14.1    & 3.05 & 0.0473/0.0572 & \\
\hline
\hline
\multicolumn{6}{|c|}{$m_{H^\pm}=m_A=800\GeV$, $m_H=600\GeV$}\\
\hline
SS1 & 39.3 & 2.14 	& 0.387 & - & -\\
SS2 & 39.3 & 1.38	 & 0.211 & - & -\\
\hline
Combined & 78.6 & 3.52  &0.598 & 0.0432/0.102 & \multirow{2}{*}{8.25/10.8}\\
\cline{1-5}
Background & 235.8 & 14.1    &3.05 & 0.0267/0.0891 & \\
\hline
\end{tabular}
\caption{Cross sections of SSDL signal for two benchmark points and  dominant SM backgrounds after different steps of  cuts. }
\label{tb:SSDL_search}
\end{table}

\begin{table}[h]
\centering
\begin{tabular}{ |c|c|c|c|c|c| } 
\hline
\hline
{\small\textbf{Combined}}	 & {\small\textbf{Total [fb]}} & {\small\pbox{15cm}{\textbf{Dilepton +1 $b$+}\\ \textbf{1 tagged top [fb]}}} & {\small\textbf{Recon. [fb]}}& {\small\pbox{15cm}{\textbf{BDT [fb]}\\ \textbf{DISC/EXC}}} &{\small$\boldsymbol{Z_{\rm{disc}}/Z_{\rm{exc}}}$}\\
\hline
\hline
\multicolumn{6}{|c|}{$m_{H^\pm}=m_A=1200\GeV$, $m_H=900\GeV$}\\
\hline
Signal & 109.7 & 7.73  &1.39 & 0.0282/0.130 & \multirow{2}{*}{4.90/6.43}\\
\cline{1-5}
Background & 707.3 & 45.7	 & 11.2 & 0.0349/0.221 &\\
\hline
\hline
\multicolumn{6}{|c|}{$m_{H^\pm}=m_A=800\GeV$, $m_H=600\GeV$}\\
\hline
Signal & 235.8 & 10.1  & 1.76 & 0.0751/0.217  & \multirow{2}{*}{7.57/10.4}\\
\cline{1-5}
Background & 707.3 & 45.7	 & 11.2 & 0.0631/0.252 &\\
\hline
\end{tabular}
\caption{The combined signal and background cross sections with cuts at two benchmark points.}
\label{tb:combined_search}
\end{table}

\subsection{Reach of Charged Higgs}

\begin{figure}[h]
\centering
 \includegraphics[width=0.49\textwidth]{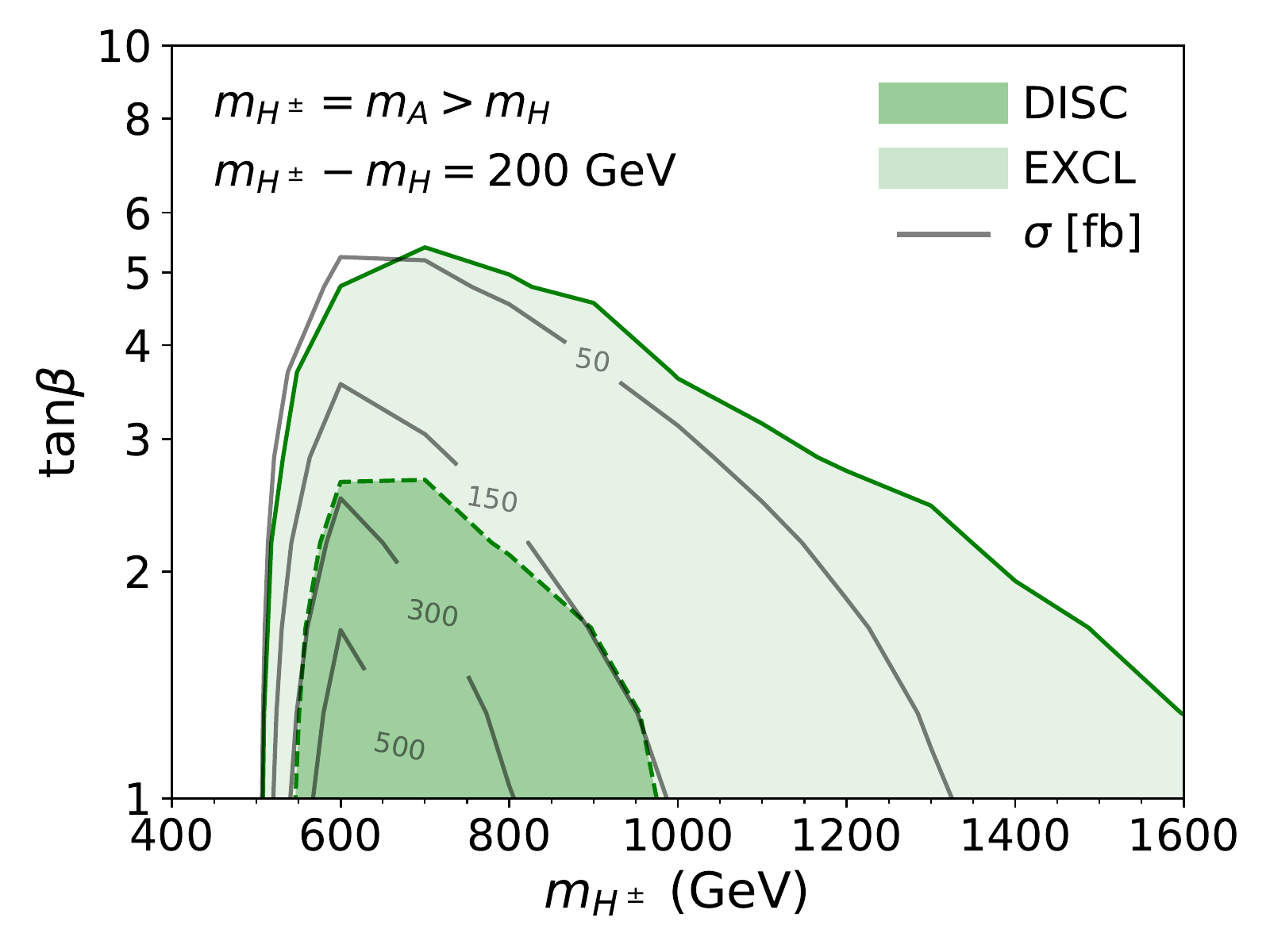}
\caption{The 95\% exclusion (region enclosed by solid green line)  and 5 $\sigma$ discovery (region enclosed by dashed green line)   reach of the exotic charged Higgs at at future 100 TeV $pp$ collider with ${\cal L}=3000\ {\rm fb}^{-1}$ through $H^\pm\to HW^\pm \to t\bar{t}W^\pm$ channel in the $m_{H^\pm}$ vs. $\tan\beta$ plane, for $m_A=m_{H^\pm}=m_H+200$ GeV. The total production cross sections for the dilepton signal are indicated by the gray curves.}
\label{fig:tanb_mC_reach}
\end{figure}

 In Fig.~\ref{fig:tanb_mC_reach}, we present the discovery (dashed lines) and exclusion (solid lines) reach
in the $m_{H^\pm}$ vs. $\tan\beta$ plane for $\bpb$ with $\mC=\mA>\mH$ at a 100 TeV $pp$ collider with  $3000~\ifb$ integrated luminosity
for a fixed mass splitting  of $\Delta m=\mC - \mH=200~\gev$.  Also shown are the production cross sections of the dilepton signal, which are indicated by the gray contour lines.

The sensitivity at large $\tan\beta$ is primarily limited by the branching fraction of $H\to t\bar t$,  which is reduced due to the competing processes $H \to b\bar{b}$ and $\tau\tau$.  For low  $\tan\beta$ around 1, this channel shows great sensitivity once it passes the $H\to t\bar t$ threshold. The advantage of top tagging is evident at large mass region, which extends the exclusion region  beyond 1.6 TeV.

\begin{figure}[h]
\centering
\includegraphics[width=0.45\textwidth]{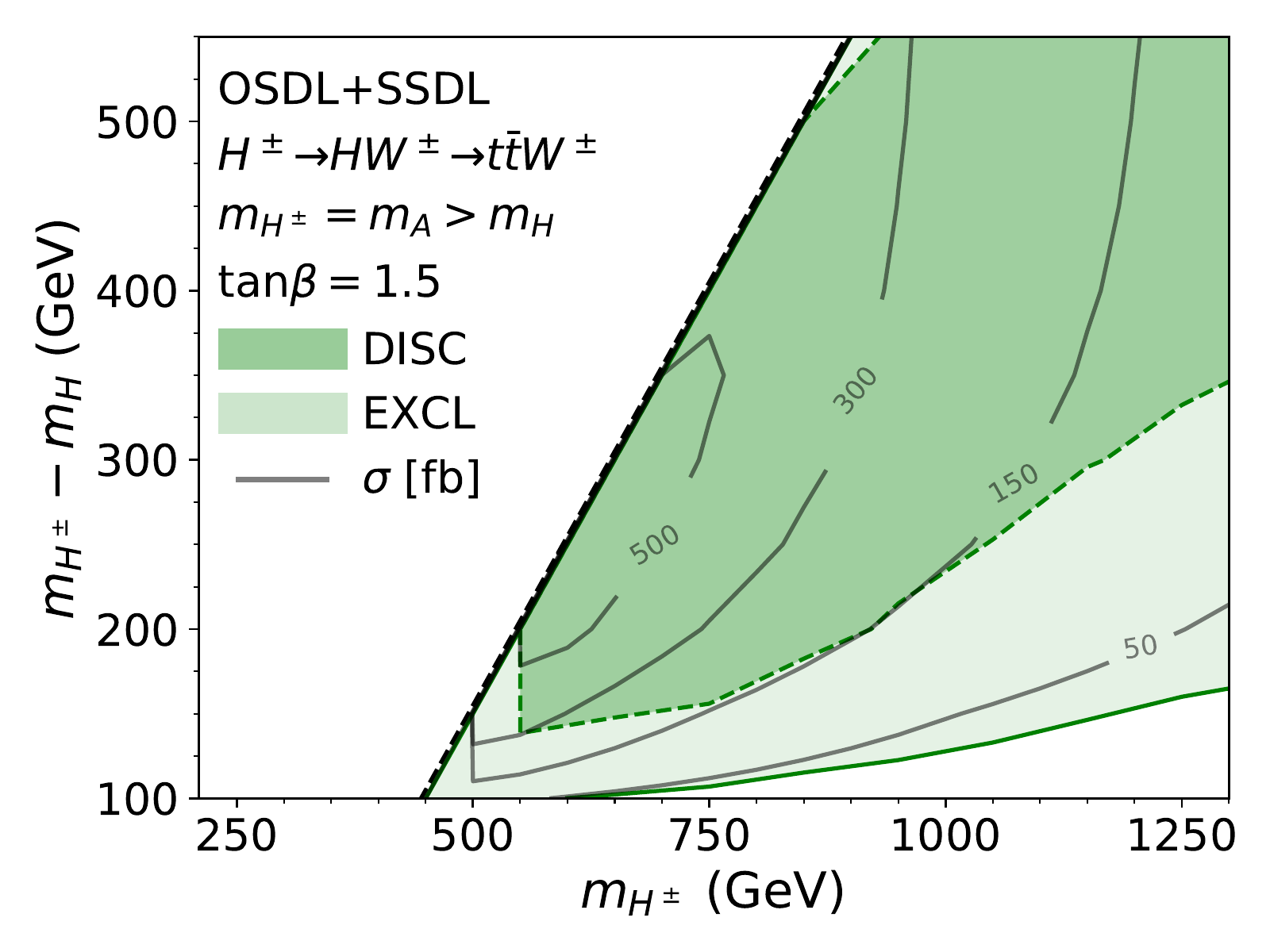}
\includegraphics[width=0.45\textwidth]{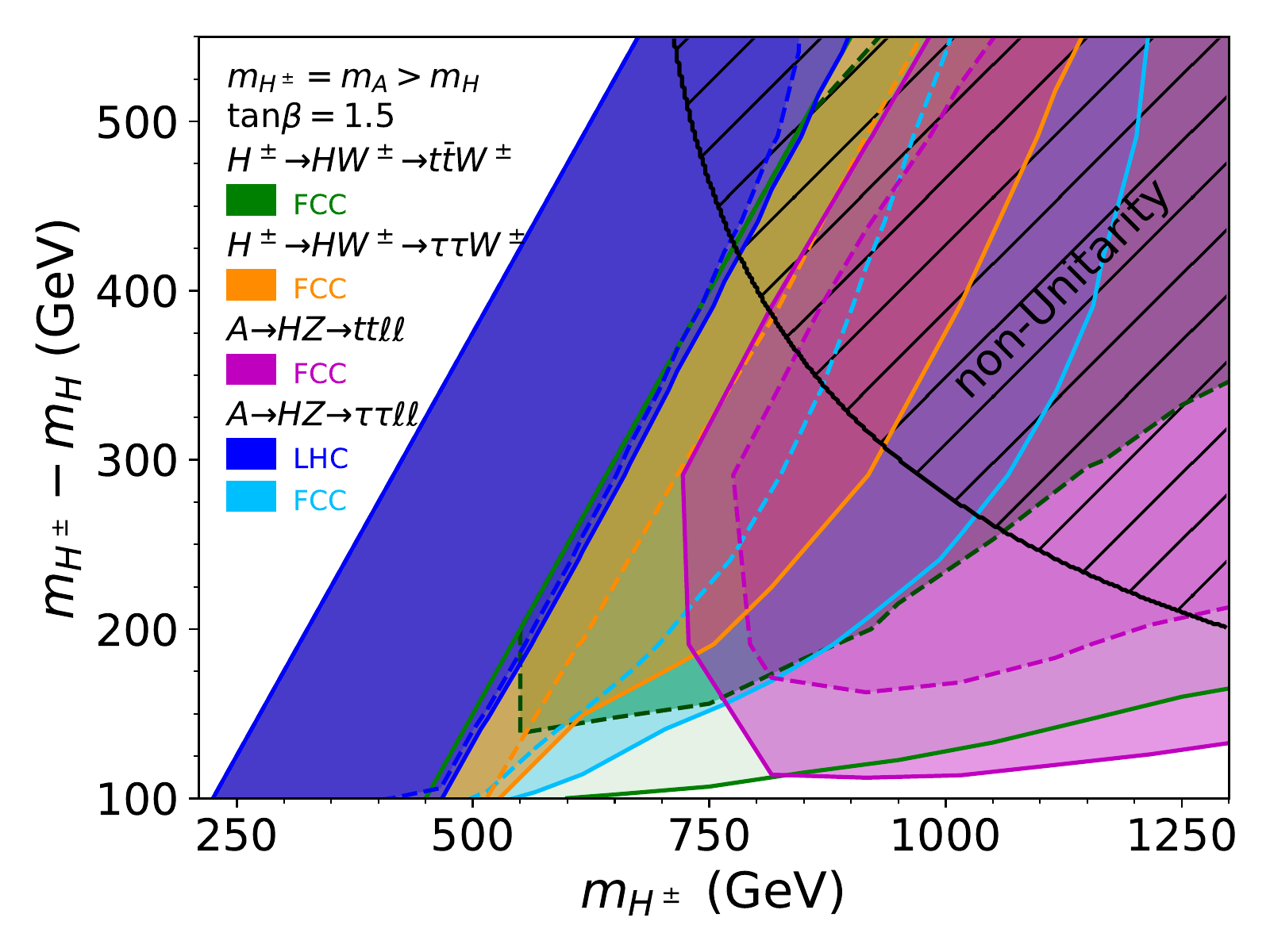}
\caption{The left panel shows the discovery and exclusion reach of charged Higgs via $H^\pm\to HW^\pm\to t\bar t W^\pm$ in $m_{H^\pm}$ vs. $m_{H^\pm}-m_H$ plane at 100 TeV $pp$ collider with ${\cal L}=3000\ {\rm fb}^{-1}$. Color coding is the same as in Fig.~\ref{fig:tanb_mC_reach}.   The right panel shows the reaches of $H^\pm\to HW^\pm\to t\bar t W^\pm$ as well as other exotic channels~\cite{Kling_2019}. See text for more detail.  }
\label{fig:dm_mC_reach}
\end{figure}

In the left panel of Fig.~\ref{fig:dm_mC_reach}, we present the reach of the charged Higgs via $H^\pm\to HW^\pm\to t\bar t W^\pm$ channel in $m_{H^\pm}$ vs. $m_{H^\pm}-m_H$ plane, for $\tan\beta=1.5$. 
  The left boundary of the enclosed region is given by the top pair mass threshold of $H$, above  which the branching fraction of $H \to t\bar{t}$   vanishes.

Consistent with Fig.~\ref{fig:tanb_mC_reach}, this channel shows great sensitivity once $m_H$ passes the top pair mass threshold. Above the top pair mass threshold, almost all mass splitting with $m_{H^\pm}-m_H\gtrsim 100\GeV$ at $m_{H^\pm}\gtrsim 500\GeV$ could be excluded at 95\% C.L.   There is no loss of sensitivity for $m_H$ near the top pair mass threshold when tops from $H$ decay are almost at rest in the rest frame of the neutral Higgs $H$.  In this parameter region, the top tagging rate for the associated hadronic tops from OS2 and SS1 signal divisions remains to be high. Meanwhile, the hadronic tops from neutral Higgs $H$ decay, despite being relatively soft, are energetic enough to be reconstructed with $p_T>100$ GeV.   

In the right panel of Fig.~\ref{fig:dm_mC_reach}, we present the reach of the  search channel in our study along with other exotic decay channels explored in Ref.~\cite{Kling_2019}.  The solid and dash lines represent the 95\% exclusion and 5 $\sigma$ discovery reaches. The hatched region is disfavored by unitarity consideration.  Except for the blue region, which shows the reach of $A \to HZ \to \tau\tau \ell\ell$ at the LHC with 300 ${\rm fb}^{-1}$ integrated luminosity, reaches of all other channels are for a 100 TeV $pp$ collider with 3000 ${\rm fb}^{-1}$ integrated luminosity.  Comparing with the gold region, which is the reach for $H^\pm\to HW^\pm\to \tau\tau W^\pm$, $H \to t\bar{t}$ channel enhance the reach above the top pair threshold with the help of top tagging technique.  This is similar to the neutral Higgs exotic decay $A \to HZ$ case, when the combination of $H \to tt$ (magenta) above the top threshold and $H \to \tau\tau$ (light blue) also covers the entire interesting parameter space.   Combing all the exotic decay channels, almost all the parameter space of \bpb~($m_{A}=m_{H^\pm}>m_{H}$) that permits the exotic decay can be explored in both the neutral and charged Higgs channels.

\section{Conclusion}
\label{sec:conclusion}
While the conventional channels for the non-SM Higgses decay into a pair of SM fermions or gauge bosons play an important role in searching for non-SM Higgses, the exotic decays of non-SM Higgses into two light non-SM Higgses or one non-SM Higgs plus one SM gauge boson dominate once they are kinematically open.  The reaches from conventional channels are relaxed under such scenario.  Exotic decays, however,  provide new opportunities for the discovery of non-SM Higgses, which are complementary to the conventional channels.  In particular, in the  subsequent decay of daughter Higgs, channels with top quarks in the final states can be studied with top tagging at 100 TeV $pp$ collider given the boosted top quark from heavy Higgs decays. 
 
In this paper, we analyzed the sensitivity of charged Higgs via the $H^{\pm}\rightarrow HW \to t\bar{t}W$ exotic decay mode at a 100 TeV $pp$ collider in a benchmark plane, \bpb~($m_{A}=m_{H^\pm}>m_{H}$), which is viable under the alignment limit after considering theoretical constraints and experimental limits.  Given the $tbH^\pm$ associated production of charged Higgses, multi-top quarks are present in the final state. A BDT classifier is trained to distinguish between the signal events and the SM background events by requiring at least one top quark being top tagged, with a second hadronic top being either top tagged or reconstructed. Two additional leptons are required to reduce the SM hadronic backgrounds.   We find that a charged Higgs with masses up to $\mC\approx 1~\tev$ at small $\tan\beta$ $\left(\approx 1\right)$ for a mass splitting of
$\mA-\mH=200~\gev$ can be discovered at 5 $\sigma$ with 3000 $\fb^{-1}$ data collected at a 100 TeV $pp$ collider.  The 95\% exclusion reach extends beyond 1.6 TeV.

We also present the combined reach in the benchmark plane \bpb{} for $\tan\beta=1.5$ in \figref{dm_mC_reach}. Decay channel $A\rightarrow HZ\rightarrow tt\ell\ell$ has already shown to be powerful to search for heavy neutral Higgs, well beyond the reach of $\tau\tau\ell\ell$ once the daughter particle $H$ is above the di-top threshold.    $H^{\pm}\rightarrow HW\rightarrow t\bar{t}W$ discussed in this paper also extends the previous  study of $H^{\pm}\rightarrow HW\rightarrow \tau\tau W$.   Combining all exotic Higgs decay channels, almost the entire parameter space in hierarchical 2HDMs can be probed at a future 100 TeV $pp$ collider.

At future energy frontier colliders, exotic decay channels provide new discovery avenues for heavy BSM  Higgses.  The discovery of an extended Higgs sector beyond the SM  could shed light on  the underlying mechanism of electroweak symmetry breaking.


\begin{acknowledgments}
We would like to thank Felix Kling, Ying-Ying Li and Tao Liu for helpful discussions.
An allocation of computer time from the  UA Research Computing High Performance Computing (HPC) and High Throughput Computing (HTC) at the University of Arizona is gratefully acknowledged.
SL, HS and SS are supported in part by the Department of Energy under Grant \texttt{DE-FG02-13ER41976/DE-SC0009913}.

\end{acknowledgments}

\bibliography{references}

\end{document}